\documentclass[aps,prb, twocolumn,superscriptaddress]{revtex4-2}
\usepackage{graphicx}
\usepackage[hidelinks]{hyperref} 
\usepackage{bm}
\usepackage{amsmath}
\usepackage{bbold}
\usepackage{xcolor}
\usepackage{float}

\graphicspath{{/DP/figs/}}

\begin{document}


\title{Dipolar and quadrupolar correlations in the $5d^2$ Re-based double perovskites Ba$_2$YReO$_6$ and Ba$_2$ScReO$_6$}

\renewcommand*{\thefootnote}{\arabic{footnote}}

\author{Otkur Omar}
\affiliation{Department of Physics, University of Science and Technology of China, Hefei, Anhui 230026, China}

\author{Yang Zhang}
\affiliation{Department of Physics and Astronomy, University of Tennessee, Knoxville, TN 37996, USA}

\author{Qiang Zhang}
\affiliation{Neutron Scattering Division, Oak Ridge National Laboratory, Oak Ridge, TN 37831, USA}

\author{Wei Tian}
\affiliation{Neutron Scattering Division, Oak Ridge National Laboratory, Oak Ridge, TN 37831, USA}

\author{Elbio Dagotto}
\affiliation{Department of Physics and Astronomy, University of Tennessee, Knoxville, TN 37996, USA}
\affiliation{Materials Science \& Technology Division, Oak Ridge National Laboratory, Oak Ridge, TN 37831, USA}

\author{Gang Chen}
\affiliation{International Center for Quantum Materials, School of Physics, Peking University, Beijing 100871, China}

\author{Taka-hisa Arima}
\affiliation{RIKEN Center for Emergent Matter Science, Wako 351-0198, Japan}
\affiliation{Department of Advanced Materials Science, University of Tokyo, Kashiwa 277-8561, Japan}

\author{Matthew B. Stone}
\affiliation{Neutron Scattering Division, Oak Ridge National Laboratory, Oak Ridge, TN 37831, USA}

\author{Andrew D. Christianson}
\affiliation{Materials Science \& Technology Division, Oak Ridge National Laboratory, Oak Ridge, TN 37831, USA}

\author{Daigorou Hirai}
\email[]{dhirai@nuap.nagoya-u.ac.jp}
\affiliation{Department of Applied Physics, Nagoya University, Nagoya 464-8603, Japan}
\affiliation{Institute of Solid State Physics, University of Tokyo, Kashiwa, Chiba 277-8581, Japan}

\author{Shang Gao}
\email[]{sgao@ustc.edu.cn}
\affiliation{Department of Physics, University of Science and Technology of China, Hefei, Anhui 230026, China}
\affiliation{Neutron Scattering Division, Oak Ridge National Laboratory, Oak Ridge, TN 37831, USA}
\affiliation{Materials Science \& Technology Division, Oak Ridge National Laboratory, Oak Ridge, TN 37831, USA}

\date{\today}

\pacs{}

\begin{abstract}
Double perovskites containing heavy transition metal ions are an important family of compounds for the study of the interplay between electron correlation and spin-orbit coupling. Here, by combining magnetic susceptibility, heat capacity, and neutron scattering measurements, we investigate the dipolar and quadrupolar correlations in two prototype rhenium-based double perovskite compounds, Ba$_2$YReO$_6$ and Ba$_2$ScReO$_6$. A type-I dipolar antiferromagnetic ground state with a propagation vector $\mathbf{q} = (0, 0, 1)$ is observed in both compounds. At temperatures above the magnetic transitions, a quadrupolar ordered phase is identified. Weak spin excitations, which are gapped at low temperatures and softened in the correlated paramagnetic phase, are explained using a minimal model that considers both the dipolar and quadrupolar interactions. At larger wavevectors, we observe dominant phonon excitations that are well described by density functional calculations.
\end{abstract}

 \maketitle

\section{Introduction}
Electron correlations and spin-orbit coupling (SOC) are two essential ingredients to achieve exotic states in condensed matter. In compounds consisting of the $4d$ or $5d$ heavy transition metal ions, the interplay between these two ingredients may lead to even richer phenomena~\cite{Witczak:annurev,chen2024multiflavor,Takayama:review}. For instance, in Sr$_2$IrO$_4$, SOC entangles the spin and orbital degrees of freedom of the Ir$^{4+}$ ($5d^5$) ions, resulting in a Mott insulating ground state formed by effective angular momentum electron bands~\cite{PhysRevLett.101.076402,SrIrO:Science}. In $\alpha$-RuCl$_3$ that is composed of the magnetic Ru$^{3+}$ ($4d^5$) ions, SOC-driven anisotropic interactions may induce a Kitaev quantum spin liquid, which holds great promise for applications in quantum computation~\cite{banerjee2016proximate,RuCl3:Science,takagi2019concept,PhysRevB.105.085107}.

\begin{figure}[htb]
    \includegraphics[width=0.42\textwidth]{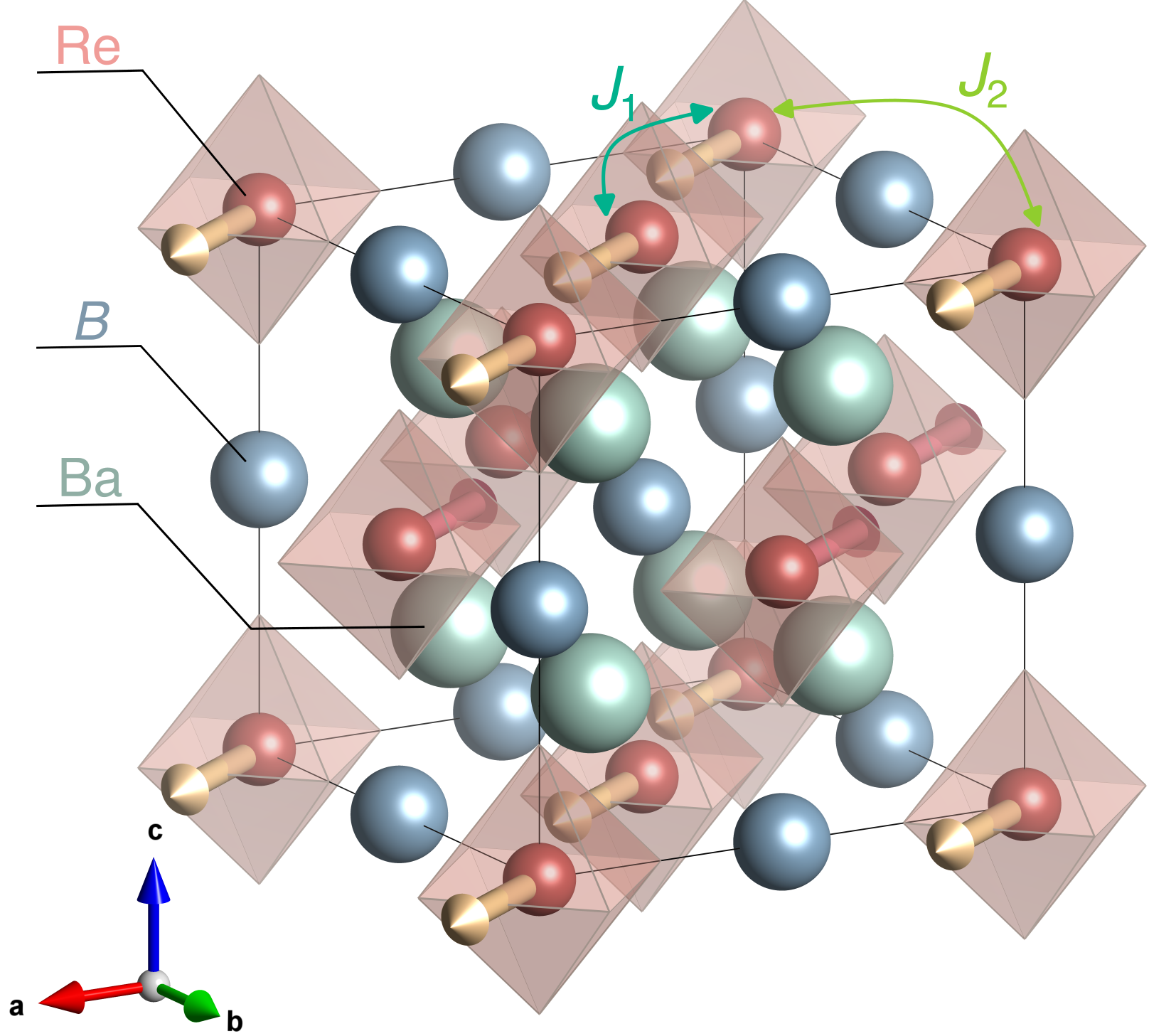}
    \caption{Crystal structure of the double perovskites Ba$_2$$B$ReO$_6$ ($B$ = Y, Sc). The Ba, $B$ = Y or Sc, and Re atoms are shown as green, blue, and red spheres, respectively. The oxygen atoms (not shown) form octahedra around the Re atoms. Exchange paths for the first-neighbor interaction, $J_1$, and second-neighbor interaction, $J_2$, are indicated by arrows. The type-I dipolar AFM order with $\mathbf{q} = (0, 0, 1)$ is indicated by yellow and pink arrows on each Re ion.}
    \label{fig:crystal}
\end{figure}

\begin{figure}[htb]
    \includegraphics[width=0.47\textwidth]{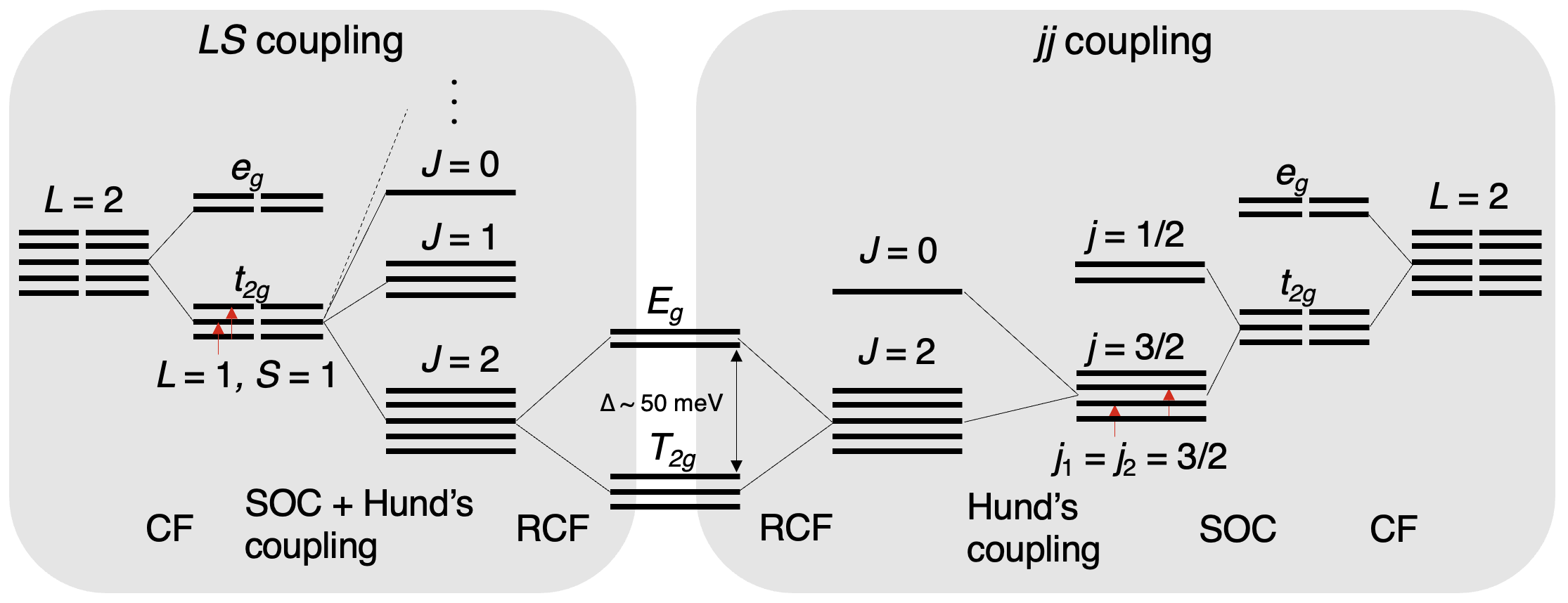}
    \caption{Schematic illustration of the energy level splitting of the $5d^2$ ion in the presence of octahedral crystal field (CF), SOC, Hund's coupling and residual crystal field (RCF). Under both the $LS$ and $jj$ coupling schemes, the ground state of the $5d^2$ ion is a dipolar and quadrupolar active triplet $T_{2g}$ manifold.
    \label{fig:GS}}
\end{figure}

The 5$d$-based double perovskites (DPs), with a chemical formula of $A_2BB'$O$_6$, are exemplary platforms for the study of the interplay between electron correlations and SOC~\cite{vasala_a2bbo6_2015}. In these materials, the $B'$ sites, being occupied by the 5$d$ heavy transition metal ions, form a face-centered cubic (FCC) lattice as shown in Fig.~\ref{fig:crystal}. Figure~\ref{fig:GS} describes the typical energy levels of the $5d^2$ heavy transition metal ions in the DPs. In these compounds, the $d$ electron orbitals with angular momentum $L = 2$ are split into a $t_{2g}$ ground state triplet and an $e_g$ excited state doublet due to a strong crystal electric field. The $t_{2g}$ ground state, which can be described as orbitals of an effective angular momentum $\tilde{l} = 1$, is further coupled to the spin degree of freedom through SOC. This leads to a ground state manifold that is described by an effective total angular moment, $J_{\mathrm{eff}}$. The energy splitting for the heavy transition metal ions in the DPs can be understood in both the $LS$ and $jj$ coupling schemes, even though the actual energy levels often lie between these two ideal scenarios due to the comparable energy scales of the Hund's coupling and SOC~\cite{Streltsov_2017,PhysRevB.95.235114,frontini2025resonantinelasticxrayscattering}. In the $5d^1$ systems, only one spin couples with the orbital, so that both coupling schemes lead to a $J_{\mathrm{eff}} = 3/2$ quartet with reduced dipolar moment sizes~\cite{erickson_ferromagnetism_2007,marjerrison_cubic_2016,hirai_successive_2019,gao_antiferromagnetic_2020,PhysRevResearch.2.022063,vzivkovic2024dynamic,hirai_muon_2020,soh_spectroscopic_2024,frontini_spin-orbit-lattice_2024,arima_interplay_2022,ishikawa_phase_2021,hirai_possible_2021}. In the $5d^3$ systems such as Ba$_2$YOsO$_6$ and Sr$_2$ScOsO$_6$, an unusually large spin gap of $\sim$ 18-19~meV has been observed in experiments~\cite{PhysRevLett.118.207202,PhysRevB.93.220408,PhysRevB.91.075133,PhysRevB.98.104434}. This excitation gap cannot be explained in the $LS$ coupling scheme as it predicts an isotropic $S = 3/2$ ground state manifold with gapless excitations. In the $jj$ coupling scheme, an excitation gap  may emerge from the anisotropic exchange interactions that are enhanced by the SOC-induced $J_{\mathrm{eff}} = 3/2$ ground state manifold~\cite{PhysRevLett.118.207202,PhysRevB.93.220408,PhysRevB.91.075133,PhysRevB.108.054436,PhysRevB.98.104434}. For the $5d^2$ compounds, both the $LS$ and $jj$ coupling schemes lead to a similar ground state manifold of $J_{\mathrm{eff}} = 2$ as illustrated in Fig.~\ref{fig:GS}. In the $LS$ coupling scheme, antisymmetrized two-electron states within the $t_{2g}$ manifold yield an effective total orbital angular momentum $L = 1$, which couples with a total spin $S = 1$ state favored by strong Hund’s coupling. Moderate SOC then mixes the $L = 1$ and $S = 1$ states to produce low-energy multiplets of effective $J_{\text{eff}} =\ 2,\ 1,\ 0$. In contrast, in the $jj$ coupling scheme, each electron first experiences strong SOC, so that the $t_{2g}$ levels are split into the $j = 1/2$ and $j = 3/2$ states. The two-electron configuration favors $j_1 = j_2 = 3/2$, leading to allowed total states of $J_{\text{eff}} = 2,\ 0$, among which the $J_{\text{eff}} = 2$ manifold is energetically favored due to Hund’s coupling. In the strong SOC limit, the magnetic moment is reduced to approximately $1.225\ \mu_\mathrm{B}$~\cite{Chen:PRB2}. For both coupling schemes, the $J_{\mathrm{eff}} = 2$ manifold can be further split by a residual crystal field (RCF) into a non-Kramers $E_g$ doublet and a $T_{2g}$ triplet~\cite{Maharaj:PhysRevLett,Paramekanti:PRB,Voleti:PRB}.

In contrast to conventional magnets with purely spin degrees of freedom, entangled spins and orbitals in the $5d$-based DPs may allow for the possibility of multipolar interactions. The study of the multipolar orders, which are often described as ‘‘hidden orders’’ due to the difficulty in their direct experimental detection, has been mainly focused on $f$-electron systems~\cite{RevModPhys.81.807,Hotta_2006}. Recent theoretical and experimental works suggest the importance of multipolar interactions in the $5d$-electron systems. A notable example is the $5d^1$ compound Ba$_2$MgReO$_6$~\cite{PhysRevResearch.2.022063,hirai_successive_2019,arima_interplay_2022,soh_spectroscopic_2024,hirai_muon_2020}. In this material, antiferroic $Q^{x^2-y^2}$ quadrupolar order and ferroic $Q^{3z^2}$ quadrupolar order have been experimentally identified. The recent discovery of the dynamic Jahn-Teller effect, driven by the orbital degree of freedom, also induced further interest in this family  of compounds~\cite{PhysRevB.98.075138, PhysRevB.107.L220404,frontini_spin-orbit-lattice_2024, vzivkovic2024dynamic,PhysRevB.110.L201101}. For the $5d^1$ and $5d^2$ systems, a rich phase diagram has been predicted by some mean-field analysis, including the type-I antiferromagnetic (AFM) dipolar order, quadrupolar order, and a variety of octupolar orders~\cite{Chen:PRB1,Chen:PRB2,Svoboda:PRB}. Recently, ferro-octupolar orders have been proposed to emerge in the Os-based $5d^2$ DPs such as Ba$_2M$OsO$_6$ ($M =$ Ca, Mg, Zn)~\cite{Maharaj:PhysRevLett}. This proposal is attributed to the absence of dipolar magnetic order and the lack of quadrupolar-related structural transitions. Many theoretical works on the $5d^2$ DPs have then been established under this framework~\cite{PhysRevB.101.054439,PhysRevB.101.155118,PhysRevB.102.064407,PhysRevB.105.014438,PhysRevLett.127.237201,banerjee:arxiv,PhysRevResearch.3.033163,PhysRevB.109.184416,kozlowski2024,PhysRevB.108.045149}. However, AFM orders, albeit with reduced ordered moment sizes, are also experimentally observed in other $5d^2$ DPs~\cite{Morrow2016,XIONG2018762,PhysRevB.103.104430}, which calls for further experimental and theoretical studies. Especially, according to the mean-field analysis of the $5d^2$ DPs~\cite{Chen:PRB2,Svoboda:PRB}, quadrupolar orders may emerge at temperatures above the magnetic dipolar ordering transition, although conclusive evidence for their existence is still missing.

Our study focuses on two archetypal $5d^2$ Re-based double perovskites, Ba$_2$YReO$_6$ and Ba$_2$ScReO$_6$. These two compounds share similar lattice geometry and electronic configurations. Previous experimental studies on Ba$_2$YReO$_6$ using muon spin rotation ($\mu$SR) and neutron powder diffraction (NPD) yield conflicting conclusions regarding to the existence of a long-range magnetic order~\cite{Thompson_2014,PhysRevB.81.064436,PhysRevB.103.104430}, while studies on Ba$_2$ScReO$_6$ have been limited to its synthesis and basic structural characterizations~\cite{sleight_compounds_1962,XXX}.

Here, through magnetic susceptibility, heat capacity, neutron diffraction, and inelastic neutron scattering (INS) experiments on polycrystalline samples of Ba$_2$YReO$_6$ and Ba$_2$ScReO$_6$, we provide experimental evidence for the existence of a type-I dipolar AFM order in both compounds. Although our INS experiments on these two compounds reveal dominant phonon excitations as confirmed through the density functional theory (DFT) calculations, weak magnetic excitations can be identified, which are gapped in the dipolar ordered phase and softened in the correlated paramagnetic phase. Under the mean-field random phase approximation (MF-RPA), a minimal model that incorporates both the dipolar and quadrupolar interactions is proposed to describe the temperature dependence of the spin dynamics, which supports the existence of a ferro-quadrupolar (FQ) order in these two compounds.

\section{Methods} 
Polycrystalline samples of Ba$_2$YReO$_6$ and Ba$_2$ScReO$_6$ were prepared by the conventional solid-state reaction~\cite{B200572G,PhysRevB.81.064436}. BaO, Y$_2$O$_3$ (Sc$_2$O$_3$), and ReO$_3$ powders were mixed in a 4:1:2 ratio in an argon-filled glove box. The mixture was pelletized and placed in a platinum capsule in an evacuated quartz tube. The tube was heated at 1000~$^{\circ}$C for 50~h and then furnace cooled. The sintered pellet was crushed and re-pelletized in a glove box, then heated in an evacuated quartz tube at 1150~$^{\circ}$C for 50~h. The obtained pellet had a dark blue color. The polycrystalline samples were characterized by X-ray diffraction (XRD) at ambient temperature in a diffractometer (SmartLab, Rigaku Corporation) with monochromatic Cu K$\alpha$ radiation.

Magnetic susceptibility was measured using a magnetic properties measurement system (MPMS3, Quantum Design). A polycrystalline pellet was attached to a quartz sample holder using varnish (GE7031, General Electric Company). Heat capacity measurements were conducted using a semi-adiabatic thermal relaxation method on a physical properties measurement system (PPMS, Quantum Design). Thermal contact between the sample and the sapphire sample stage was made using grease (Apiezon-N, M \& I Materials Ltd). 

Neutron diffraction experiments on powder samples of Ba$_2$YReO$_6$ and Ba$_2$ScReO$_6$ were performed on the POWGEN diffractometer~\cite{POWGEN} at the Spallation Neutron Source (SNS) of the Oak Ridge National Laboratory (ORNL) and the VERITAS (HB-1A) triple-axis spectrometer at the High Flux Isotope Reactor (HFIR) of the ORNL. For the experiments on POWGEN, about 8.5~g powder of Ba$_2$YReO$_6$ and Ba$_2$ScReO$_6$ were filled into air-tight vanadium cans with a diameter of 8~mm in a helium-filled glovebox. The Powgen Automatic Changer was utilized to reach a base temperature of $\sim18$~K. The neutron frame 2 with center-wavelength of 1.5~\AA~was used to collect the data, covering a wide $Q$ region from 0.48~\AA$^{-1}$ to 12.98~\AA$^{-1}$. Data reduction was performed using the Mantid software~\cite{MANTID}. For the experiments on the HB-1A instrument, the same powder samples were filled in vanadium sample cans with a diameter of 17~mm in a helium-filled glovebox. An incoming neutron wavelength of $\lambda = 2.38$~\AA~was selected using a PG(002) monochromator. A PG(002) analyzer was employed to ensure elastic scattering, leading to an energy resolution of 1~meV. A closed cycle refrigerator (CCR) cooling machine was utilized to reach the base temperature of 4~K.

Inelastic neutron scattering (INS) experiments on powder samples of Ba$_2$YReO$_6$ and Ba$_2$ScReO$_6$ were performed on the SEQUOIA spectrometer~\cite{SEQUOIA} at the SNS. The same powder samples used for diffraction measurements were sealed in aluminum
cans in a helium-filled glovebox. A CCR cooling machine was utilized to reach a base temperature
of 6~K. Measurements were taken with incident neutron energies of $E_i = 60$~meV in the high flux chopper configuration with a Fermi chopper frequency of 120~Hz and $E_i = 225$~meV in the high resolution chopper configuration with a Fermi chopper frequency of 600~Hz. For both measuring conditions, data were also collected using an empty can and subtracted as background. Data were histogrammed using the Mantid software~\cite{MANTID} and further data reduction was performed using the Mslice program in DAVE~\cite{DAVE}.

To understand the experimental results, we used the first-principles DFT technique as implemented on the Vienna {\it ab initio} simulation package (VASP) package with the projector augmented wave (PAW) method~\cite{Kresse:Prb,Kresse:Prb96,Blochl:Prb}. Electronic correlations were considered by using the generalized gradient approximation (GGA) with the Perdew-Burke-Ernzerhof (PBE) potential~\cite{Perdew:Prl}. Furthermore, the atomic positions were fully relaxed until the Hellman-Feynman force on each atom was smaller than $0.001$~eV/{\AA} using the A-type magnetic state obtained in our neutron experiment, where the lattice constants were fixed as collected in the diffraction data. In addition, the on-site interactions were considered by using the local spin-density approach (LSDA) plus $U_{\rm eff}$ ($U_{\rm eff} = 2$~eV) with the Dudarev's rotationally invariant formulation~\cite{Dudarev:PrB}. Then, we calculated the force constants by using the density functional perturbation theory approach~\cite{Baroni:Prl,Gonze:Pra1}, and analyzed the phonon dispersion relations by the PHONONPY software~\cite{Chaput:prb,Togo:sm}. Here, the plane-wave cutoff energy was set as $600$~eV and the $k$-point mesh was used as $4\times4\times4$ for the conventional cell for both Ba$_2$YReO$_6$ and Ba$_2$ScReO$_6$. The calculated data are convoluted with the instrumental energy resolution using the OCLIMAX program~\cite{OCLIMAX}.

To analyze the INS spectra and the temperature evolution of the magnetic susceptibility, we employed the mean-field random phase approximation (MF-RPA)~\cite{Jensen:RareEarth,Rotter_2012,Boothroyd_neutronbook}. This method allows for the calculation of spin excitations by treating the dipolar and quadrupolar interactions on an equal basis. The magnetic susceptibility calculations were performed with an energy step of 0.02~meV. For the calculation of the INS cross section, an energy step of 0.3~meV was used. The powder average of the calculated INS spectra was obtained by sampling 100 random points at each $Q$ position in steps of 0.02~\AA$^{-1}$. The resulting spectra were convoluted by the instrumental energy resolution function.
\section{Experimental Results} 
\subsection{Crystal Structure}
\label{sec:crystal}
\begin{figure}[htb]
    \includegraphics[width=0.47\textwidth]{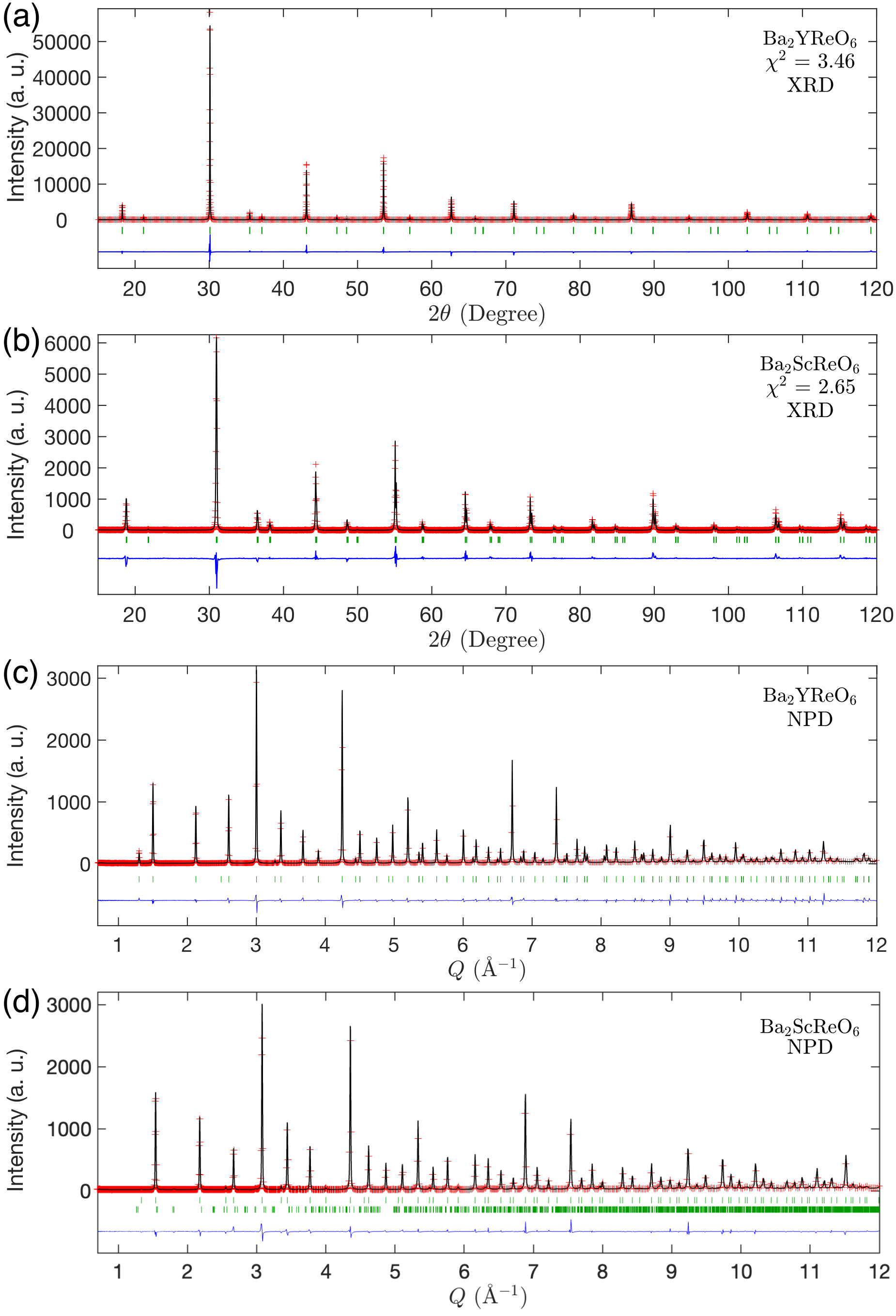}
    \caption{Refinement results of the X-ray diffraction data for
(a) Ba$_2$YReO$_6$ and (b) Ba$_2$ScReO$_6$ at $T$ = 300~K. Refinement results of the neutron diffraction data for (c) Ba$_2$YReO$_6$ and (d) Ba$_2$ScReO$_6$ at room temperature. Red crosses represent the observed data points, the solid line corresponds to the calculated pattern, and the blue line shows the difference between the observed and calculated patterns. The green ticks indicate the positions of nuclear Bragg reflections. For Ba$_2$ScReO$_6$, the lower row of green ticks in panel (d) indicates the presence of tiny Sc$_{2+x}$O$_{3+y}$ impurity, for which the diffraction intensity was calculated by the le Bail method.}
    \label{fig:XRD}
\end{figure}

Figure~\ref{fig:XRD} presents the Rietveld refinement results of the XRD patterns collected at room temperature, along with the neutron diffraction patterns collected on POWGEN at $T$ = 300~K. Table \ref{table:BYRO} and \ref{table:BSRO} summarize the refined parameters of the neutron diffraction pattern for Ba$_2$YReO$_6$ and Ba$_2$ScReO$_6$, respectively. A tiny amount of Sc$_{2+x}$O$_{3+y}$ impurity phase with an estimated mass fraction of less than $\sim0.02$~wt$\%$ was identified in Ba$_2$ScReO$_6$, while no impurity phase is detectable in Ba$_2$YReO$_6$. Throughout the entire temperature regime investigated, both Ba$_2$YReO$_6$ and Ba$_2$ScReO$_6$ stay in the cubic space group $Fm\bar{3}m$ as consistent with the previous report \cite{PhysRevB.81.064436,sleight_compounds_1962}. For the refinement of the neutron diffraction patterns, introducing off-stoichiometric occupancy does not improve the goodness-of-fit for Ba$_2$YReO$_6$. While for Ba$_2$ScReO$_6$, the $R$-factor decreases from  5.99~$\%$ to 5.57~$\%$ when the occupancy of the Sc and Re sites is fitted to 0.902(8) and 1.05(1). This defect in Ba$_2$ScReO$_6$ can be attributed to relatively closer radii of the Sc$^{3+}$ and Re$^{5+}$ ions, which puts Ba$_2$ScReO$_6$ close to the stability boundary of the ordered double perovskites~\cite{ANDERSON1993197, vasala_a2bbo6_2015}.
\begin{table*}[htb]
\centering
\caption{The refined cell parameters and atomic positions of Ba$_2$YReO$_6$ using the diffraction data collected on POWGEN at $T$ = 300 and 18~K.}
\begin{tabular}{lcccccccc}
\hline\hline
 & \multicolumn{4}{c}{300~K} & \multicolumn{4}{c}{18~K} \\
\cline{2-5} \cline{6-9}
Atom & $x$ & $y$ & $z$ & $U_{\text{iso}}$ (\AA$^2$) & $x$ & $y$ & $z$ & $U_{\text{iso}}$ (\AA$^2$) \\
\hline  
Ba & 0.25 & 0.25 & 0.25 & 0.0060(2) & 0.25 & 0.25 & 0.25 & 0.0007(1) \\
Y  & 0.5  & 0.5  & 0.5  & 0.0053(3) & 0.5  & 0.5  & 0.5  & 0.0006(2) \\
Re & 0    & 0    & 0    & 0 & 0    & 0    & 0    & 0 \\
O  & 0.23427(9) & 0 & 0 & 0.0087(2) & 0.23427(6) & 0 & 0 & 0.0036(1) \\
\hline
$a$ & \multicolumn{4}{c}{8.37091(2)} & \multicolumn{4}{c}{8.36033(2)} \\
$R_p$ & \multicolumn{4}{c}{7.67$\%$} & \multicolumn{4}{c}{8.52$\%$} \\
$R_{wp}$ & \multicolumn{4}{c}{10.50$\%$} & \multicolumn{4}{c}{11.72$\%$} \\
\hline\hline
\end{tabular}
\label{table:BYRO}
\end{table*}
\begin{table*}[ht]
\centering
\caption{The refined cell parameters and atomic positions of Ba$_2$ScReO$_6$ using the diffraction data collected on POWGEN at $T$ = 300 and 18~K.}
\begin{tabular}{lcccccccc}
\hline\hline
 & \multicolumn{4}{c}{300~K} & \multicolumn{4}{c}{18~K} \\
\cline{2-5} \cline{6-9}
Atom & $x$ & $y$ & $z$ & $U_{\text{iso}}$ (\AA$^2$) & $x$ & $y$ & $z$ & $U_{\text{iso}}$ (\AA$^2$) \\
\hline
Ba & 0.25 & 0.25 & 0.25 & 0.0053(2) & 0.25 & 0.25 & 0.25 & 0.0033(1) \\
Sc  & 0.5  & 0.5  & 0.5  & 0.0039(4) & 0.5  & 0.5  & 0.5  & 0.0061(2) \\
Re & 0    & 0    & 0    & 0.0047(5) & 0    & 0    & 0    & 0 \\
O  & 0.2429(2) & 0 & 0 &  0.0083(1) & 0.2425(1) & 0 & 0 & 0.0069(1) \\
\hline
$\text{Occ. (Sc)}$ & \multicolumn{4}{c}{0.902(8)} & \multicolumn{4}{c}{--} \\
$\text{Occ. (Re)}$ & \multicolumn{4}{c}{1.05(1)} & \multicolumn{4}{c}{--} \\
$a$ & \multicolumn{4}{c}{8.16177(2)} & \multicolumn{4}{c}{8.15103(2)} \\
$R_p$ & \multicolumn{4}{c}{5.72$\%$} & \multicolumn{4}{c}{7.04$\%$} \\
$R_{wp}$ & \multicolumn{4}{c}{7.97$\%$} & \multicolumn{4}{c}{8.04$\%$} \\
\hline\hline
\end{tabular}
\label{table:BSRO}
\end{table*}
\subsection{Magnetization}
\begin{figure}[htb]
    \includegraphics[width=0.47\textwidth]{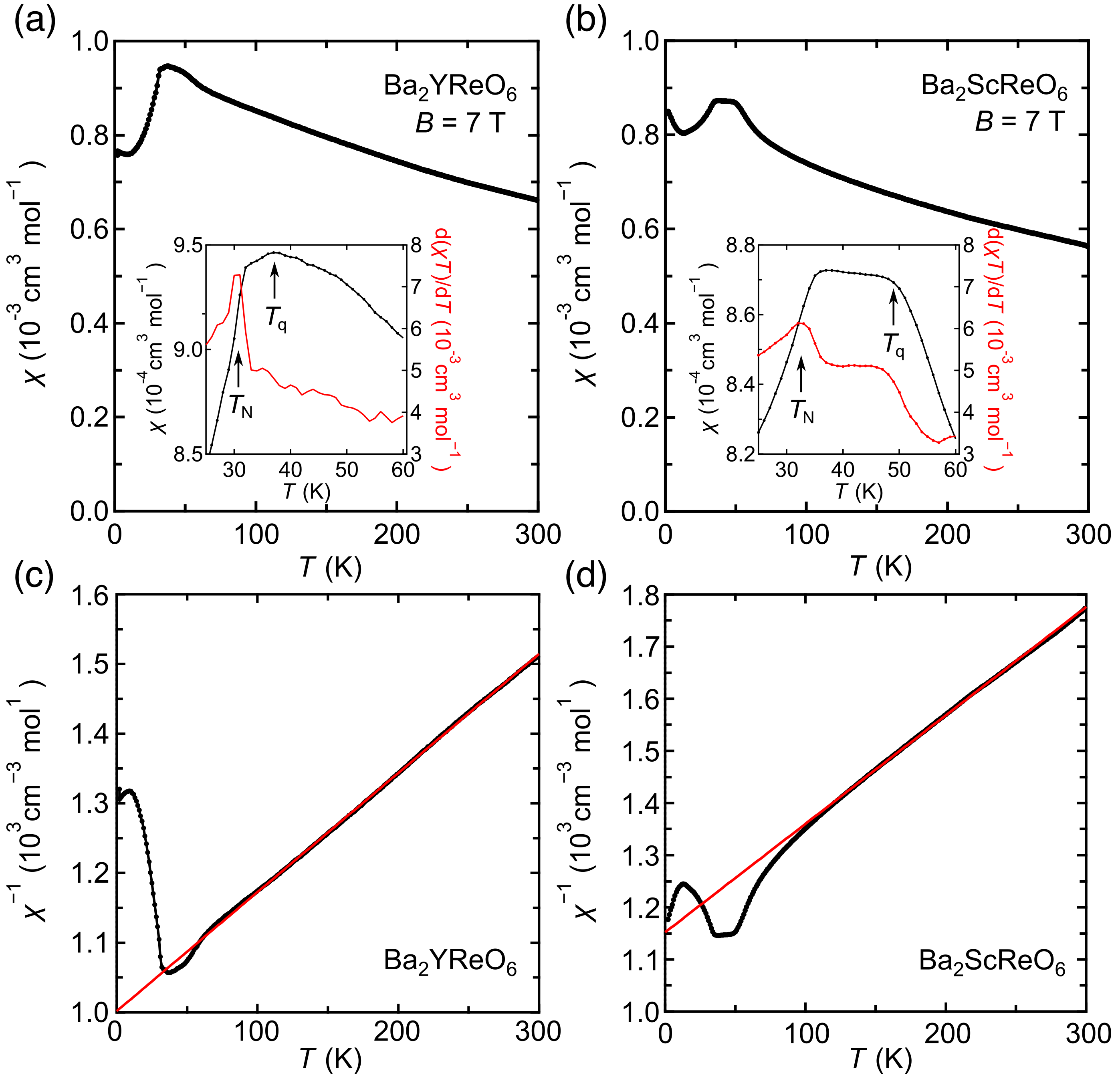}
    \caption{Temperature evolution of the magnetic susceptibility, $\chi(T)$, for (a) Ba$_2$YReO$_6$ and (b) Ba$_2$ScReO$_6$ measured in a 7~T field. Inset is an enlarged view of the low temperature regime with red solid line corresponding to the Fisher heat capacity $\mathrm{d}(\chi T)/\mathrm{d}T$. Inverse magnetic susceptibility, $\chi^{-1}(T)$, for (c) Ba$_2$YReO$_6$ and (d) Ba$_2$ScReO$_6$. The red solid line is the Curie-Weiss fit for the magnetic susceptibility in the high temperature regime of 150 - 300~K.}
    \label{fig:chi}
\end{figure}

Figures~\ref{fig:chi}(a) and~\ref{fig:chi}(b) summarize the temperature dependence of the magnetic susceptibility, $\chi$($T$), for Ba$_2$YReO$_6$ and Ba$_2$ScReO$_6$ measured in a field of 7~T. For Ba$_2$YReO$_6$, a sudden drop in $\chi$($T$) is observed at $T_{\mathrm{N}}$ = 31~K, which suggests the existence of an AFM long-range ordering transition at this temperature. At a slightly higher temperature of $T_{\mathrm{q}}\sim$ 37~K, an additional transition can be identified as the maximum of the $\chi$($T$) value. Similarly, in Ba$_2$ScReO$_6$, two transitions are observed at $T_{\mathrm{N}}=35$~K and $T_{\mathrm{q}}\sim 50$~K. In the intermediate temperature range between $T_{\mathrm{N}}$ and $T_{\mathrm{q}}$, a plateau is observed, which is more pronounced in Ba$_2$ScReO$_6$. To better elucidate the magnetic transition and compare with the heat capacity data in the next section, we plotted the $\mathrm{d}(\chi T)/\mathrm{d}T$, i.e. the Fisher relation \cite{Fisher01101962}, in the inset of the Figs.~\ref{fig:chi}(a) and~\ref{fig:chi}(b). It reveals a clear maximum for  Ba$_2$YReO$_6$ at $T_{\mathrm{N}}=31$~K, while for Ba$_2$ScReO$_6$, a relatively broader peak with a maximum at $T_{\mathrm{N}}=32.5$~K is observed. According to mean-field calculations, the two successive transitions at $T_{\mathrm{N}}$ and $T_{\mathrm{q}}$ may correspond to the magnetic dipolar and quadrupolar long-range order transitions, respectively~\cite{Chen:PRB2,Svoboda:PRB}. Figures~\ref{fig:chi}(c) and~\ref{fig:chi}(d) present the inverse magnetic susceptibility data for Ba$_2$YReO$_6$ and Ba$_2$ScReO$_6$. The red solid line is the Curie-Weiss fit of the magnetic susceptibility in the high-temperature regime of 150 - 300~K. At temperatures above 150~K, both DPs display well-defined Curie-Weiss behavior, yielding an effective magnetic moment $\mu_{\mathrm{eff}} = 2.16\  \mu_{\mathrm{B}}$, and Weiss temperature $\Theta_\mathrm{W}$ = $-$584(1)~K for Ba$_2$YReO$_6$, and $\mu_{\mathrm{eff}} = 1.98\ \mu_{\mathrm{B}}$ and $\Theta_\mathrm{W}$ = $-$570(1)~K for Ba$_2$ScReO$_6$. At $T \sim$ 60~K for Ba$_2$YReO$_6$ and $T \sim$ 100~K for Ba$_2$ScReO$_6$, a deviation from the Curie-Weiss behavior is observed, which suggests the development of short-range fluctuations. A similar phenomenon has also been observed in the previous report for Ba$_2$YReO$_6$ with fitted $\mu_{\mathrm{eff}} = 1.93\  \mu_{\mathrm{B}}$ and $\Theta_\mathrm{W}$ = $-616(7)$~K~\cite{PhysRevB.81.064436}. The reduced effective moment size, which is lower than the spin-only value of $2.83\ \mu_{\mathrm{B}}$, indicates a significant compensation from the orbital moment. It is noteworthy that the Weiss temperatures for Ba$_2$YReO$_6$ and Ba$_2$ScReO$_6$ are approximately four times higher than that of Ba$_2$CaOsO$_6$, a closely related $5d^2$ compound that does not exhibit magnetic dipolar order at low temperatures~\cite{Maharaj:PhysRevLett}. 
\begin{figure}[htb]
    \includegraphics[width=0.47\textwidth]{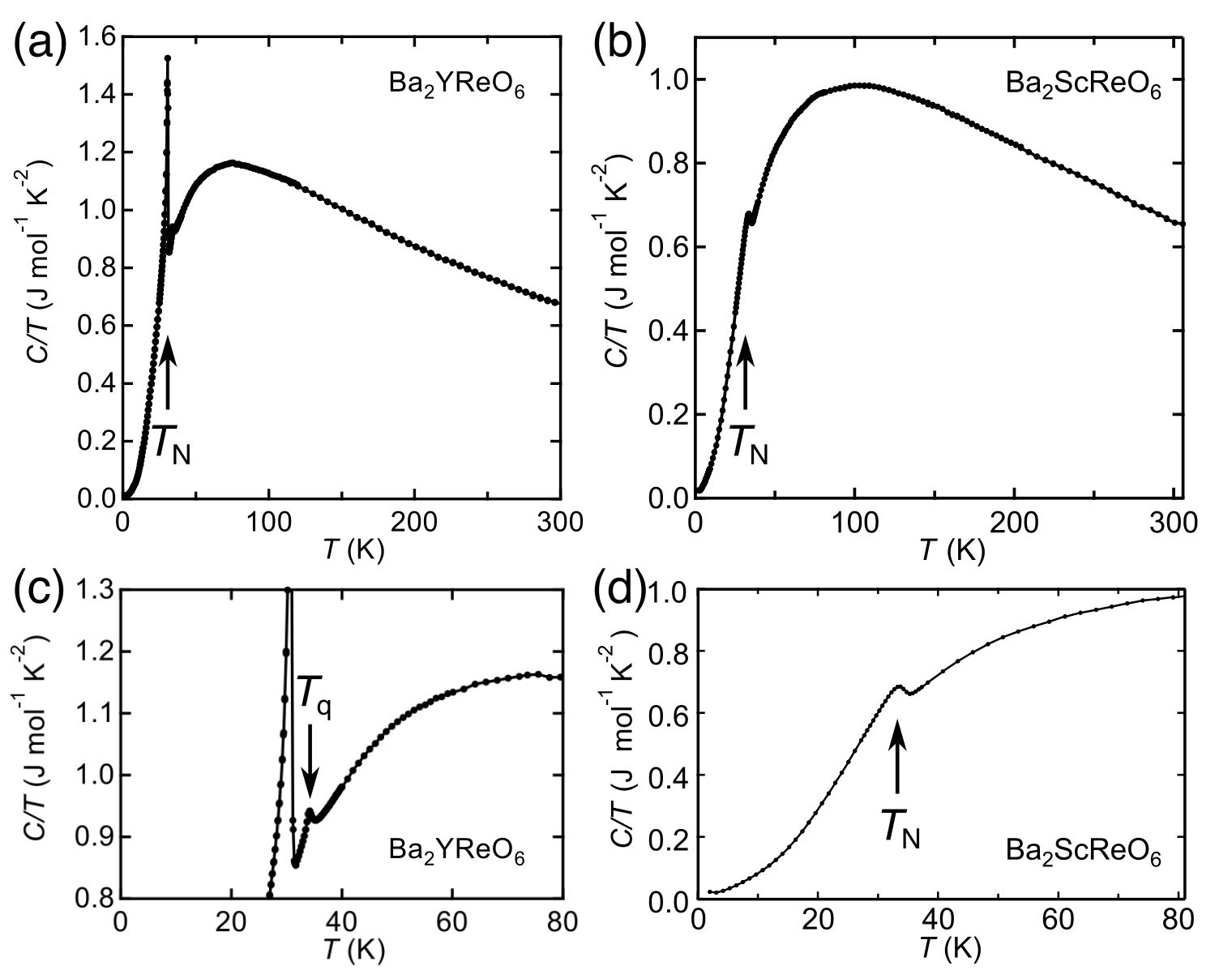}
    \caption{ Temperature evolution of the normalized heat capacity, $C(T)/T$, for (a) Ba$_2$YReO$_6$ and (b) Ba$_2$ScReO$_6$. A zoomed-in view of the data in the low temperature regime is shown in (c) Ba$_2$YReO$_6$ and (d) Ba$_2$ScReO$_6$.
    \label{fig:Cp}}
\end{figure}
\begin{figure}[htb]
    \includegraphics[width=0.47\textwidth]{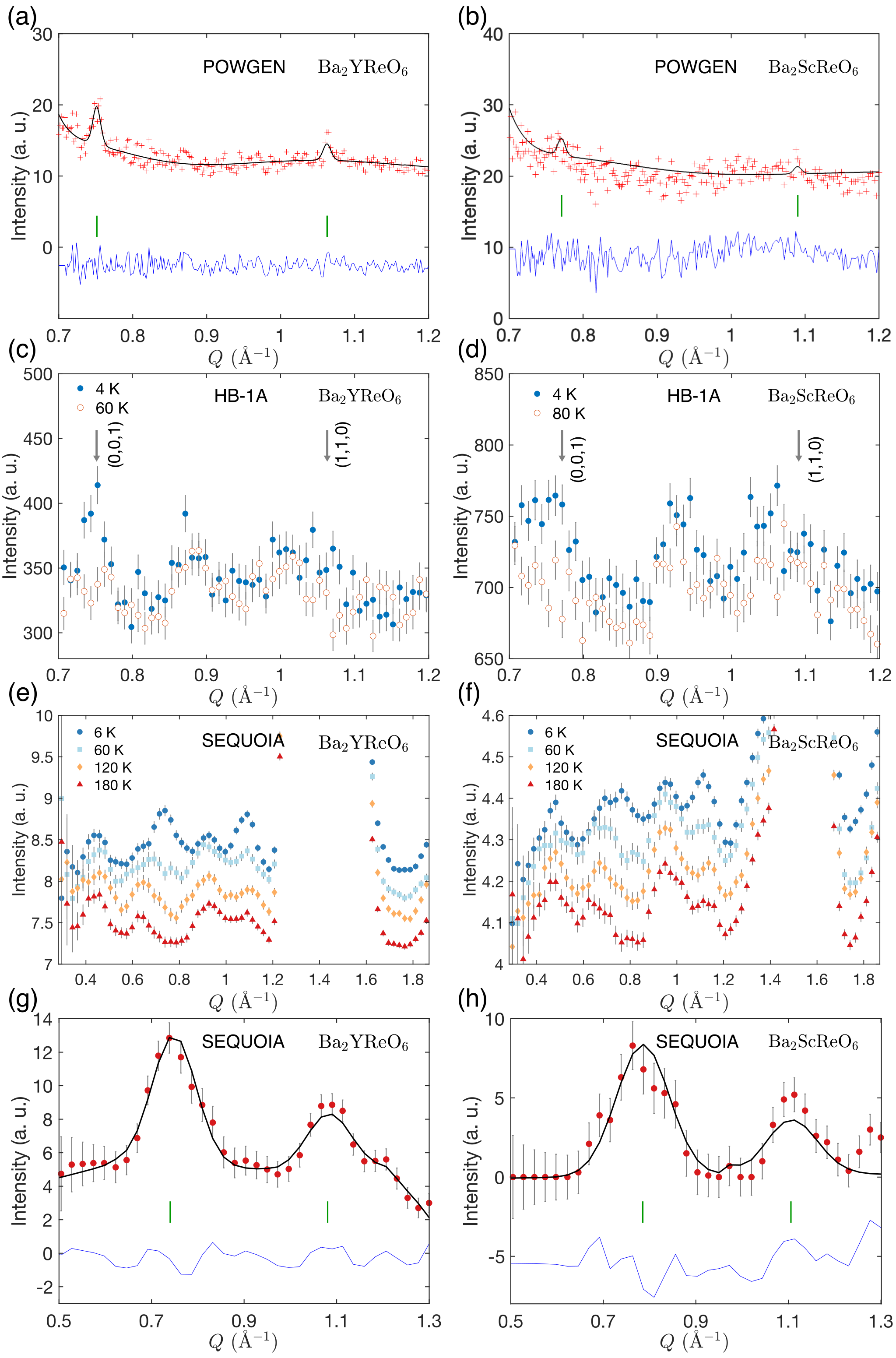}
    \caption{(a, b) Refinement results of the NPD data collected on POWGEN at $T$ = 18~K for (a) Ba$_2$YReO$_6$ and (b) Ba$_2$ScReO$_6$. Symbols in the figure are the same in Fig.~\ref{fig:XRD}. Green ticks highlight the magnetic Bragg peaks at $(0, 0, 1)$ and $(1, 1, 0)$. (c, d) NPD data collected on HB-1A at $T$ = 4, 60, 80~K for Ba$_2$YReO$_6$ and Ba$_2$ScReO$_6$, respectively. (e, f) Temperature dependence of the elastic signals obtained by integrating the $E_{\mathrm{i}}$ = 60 meV SEQUOIA data in an energy transfer range of [$-1.5, 1.5$] meV at $T$ = 6, 60, 120, 180~K. For Ba$_2$YReO$_6$, data are offset by 0, 0.1, 0.2, and 0.3 at 6, 60, 120, and 180~K, respectively. For Ba$_2$ScReO$_6$, data are offset by 0, 0.05, 0.1, and 0.2 at 6, 60, 120, and 180~K, respectively. (g, h) Rietveld refinements were performed on the difference data obtained by subtracting the 180~K data from the 6~K data on SEQUOIA for both DPs. The red points represent the observed data, the black solid line corresponds to the calculated pattern, and the blue line below shows the difference between the observed and calculated patterns. The vertical green ticks indicate the positions of magnetic peaks.
    \label{fig:NPD}}
\end{figure}
\begin{figure}[htb]
    \includegraphics[width=0.47\textwidth]{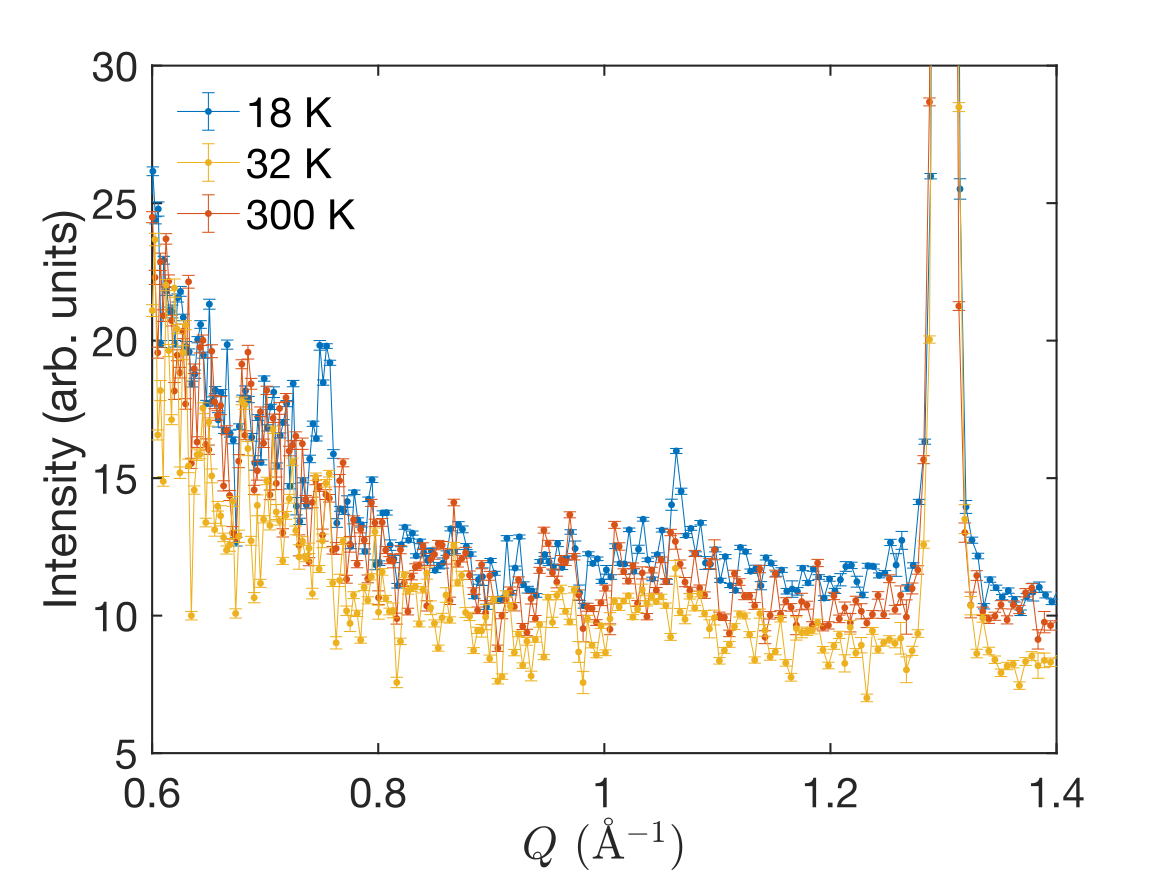}
    \caption{Temperature dependence of the neutron diffraction data for Ba$_2$YReO$_6$ collected on POWGEN at $T=$ 18, 32, and 300~K. Error bars are shown but are smaller than symbol sizes for most data in the figure.
    \label{fig:POWGEN_Tdep}}
\end{figure}
\subsection{Heat Capacity}
Figure~\ref{fig:Cp} presents the temperature evolution of the normalized heat capacity, $C(T)/T$, for Ba$_2$YReO$_6$ and Ba$_2$ScReO$_6$. The transition temperature for both compounds is consistent with the magnetization measurements. For Ba$_2$YReO$_6$, $C(T)/T$ exhibits a sharp $\lambda$-shaped peak at $T_{\mathrm{N}}=31$~K, which is not observed in the previous report~\cite{B200572G,PhysRevB.81.064436}. This difference may arise from the improved quality of our synthesized sample: In our experiments, we observe that a higher sintering temperature improves the crystallization and leads to a more pronounced transition in $C(T)$. An additional transition at $\sim 35\ \mathrm{K}$ is observed for Ba$_2$YReO$_6$ in Fig.~\ref{fig:Cp}(c), which is close to the transition at $T_{\mathrm{q}}=37$~K in $\chi$($T$). For Ba$_2$ScReO$_6$, a relatively weak peak is observed at 33.5~K, which is close to the AFM transition at $T_{\mathrm{N}}=35$~K in $\chi$($T$) shown in Fig.~\ref{fig:chi}(b). Further attempts to improve the crystallization of the Ba$_2$ScReO$_6$ sample through increased sintering temperature were not successful. As discussed in Section~\ref{sec:crystal}, Ba$_2$ScReO$_6$ lies near the stability boundary of rocksalt ordered phase, making it challenging to obtain high-quality samples.


\subsection{Magnetic Order}
Neutron diffraction data for powder samples of Ba$_2$YReO$_6$ and Ba$_2$ScReO$_6$ were collected on POWGEN at $T$ = 18, 32, and 300~K. As shown in Figs.~\ref{fig:NPD}(a) and~\ref{fig:NPD}(b), at $T$ = 18~K, magnetic Bragg peaks belonging to the propagation vector $\mathbf{q} = (0, 0, 1)$ are observed. For both samples, the magnetic structures are refined to be a type-I dipolar AFM order as described in Fig.~\ref{fig:crystal}. The magnetic form factor of the Re$^{5+}$ ions as reported in Ref.~\cite{Kobayashi:kx5002} is employed for our refinements. For Ba$_2$YReO$_6$ and Ba$_2$ScReO$_6$, the ordered dipolar moment sizes are refined to be $0.41(4) \ \mu_{\mathrm{B}}$ and $0.3(2)\ \mu_{\mathrm{B}}$, respectively. These refined moment sizes are consistent with the refined value in the previous polarized NPD experiment for Ba$_2$YReO$_6$ \cite{PhysRevB.103.104430}, and are also close to the predicted values in the DFT calculations \cite{PhysRevB.97.235119,KUKUSTA2025172714}. The reduced ordered  moment size in Ba$_2$YReO$_6$ and Ba$_2$ScReO$_6$ may result from a competition between the dipolar and quadrupolar orders~\cite{PhysRevB.103.104429,PhysRevB.103.104430}, which is also discussed in the following MF-RPA analysis section. Neutron diffraction data were also collected on HB-1A at $T$ = 4, 60, 80~K as shown in Figs.~\ref{fig:NPD}(c) and~\ref{fig:NPD}(d), which reveal the same magnetic Bragg peaks from the POWGEN experiments. As further evidence for the existence of a dipolar long-range order at low temperatures, Figs.~\ref{fig:NPD}(e) and~\ref{fig:NPD}(f) present the elastic signals obtained by integrating the SEQUOIA data in the energy transfer range of [$-1.5, 1.5$]~meV. At $T$ = 6~K, magnetic peaks of $(0, 0, 1)$ and $(1, 1, 0)$ are observed for both samples. Magnetic structural refinements were performed for the data collected at $T$ = 6~K with the data collected at $T$ = 180~K being subtracted as the background. The refinement results are summarized in Figs.~\ref{fig:NPD}(g) and~\ref{fig:NPD}(h). The refined ordered moment sizes are $0.47(1)$ and $0.52(2)\ \mu_{\mathrm{B}}$ for Ba$_2$YReO$_6$ and Ba$_2$ScReO$_6$, respectively. The difference in the ordered moment size compared to the values refined from the POWGEN data is likely due to their different measuring temperatures. Similar magnetic diffraction patterns have also been observed in other $5d^2$ DPs~\cite{XIONG2018762,Morrow2016}. Meanwhile, at elevated temperatures of 32 and 300~K, as shown in Fig.~\ref{fig:POWGEN_Tdep}, the magnetic Bragg peaks disappear, which indicates the transition at 35~K in the specific heat data is not of a dipolar origin.

\begin{figure*}[htb]
    \includegraphics[width=1.0\textwidth]{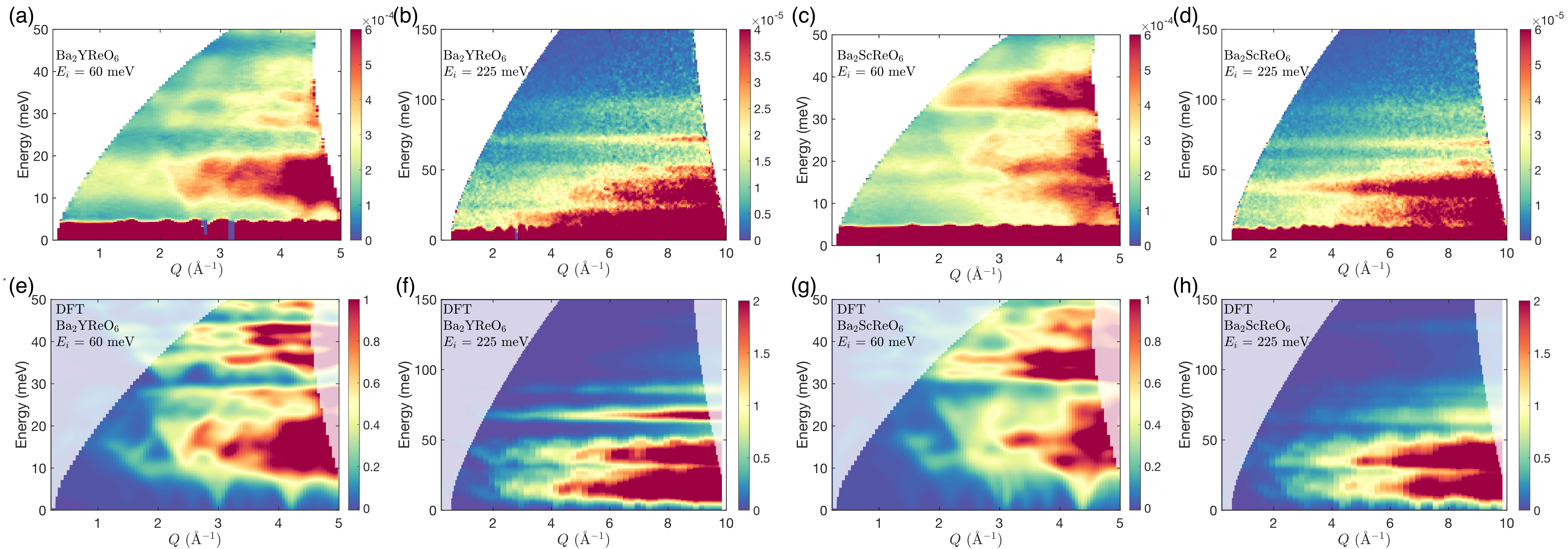}
    \caption{(a-d) INS data collected on SEQUOIA at $T$ = 6~K for Ba$_2$YReO$_6$ and Ba$_2$ScReO$_6$ with incident energies $E_\mathrm{i}$ = 60~meV and 225~meV. (e-h) Calculated phonon spectra convolved by the corresponding instrumental energy resolution functions.
    \label{fig:DFT}}
\end{figure*}
\begin{figure*}[htb]
    \includegraphics[width=1.0\textwidth]{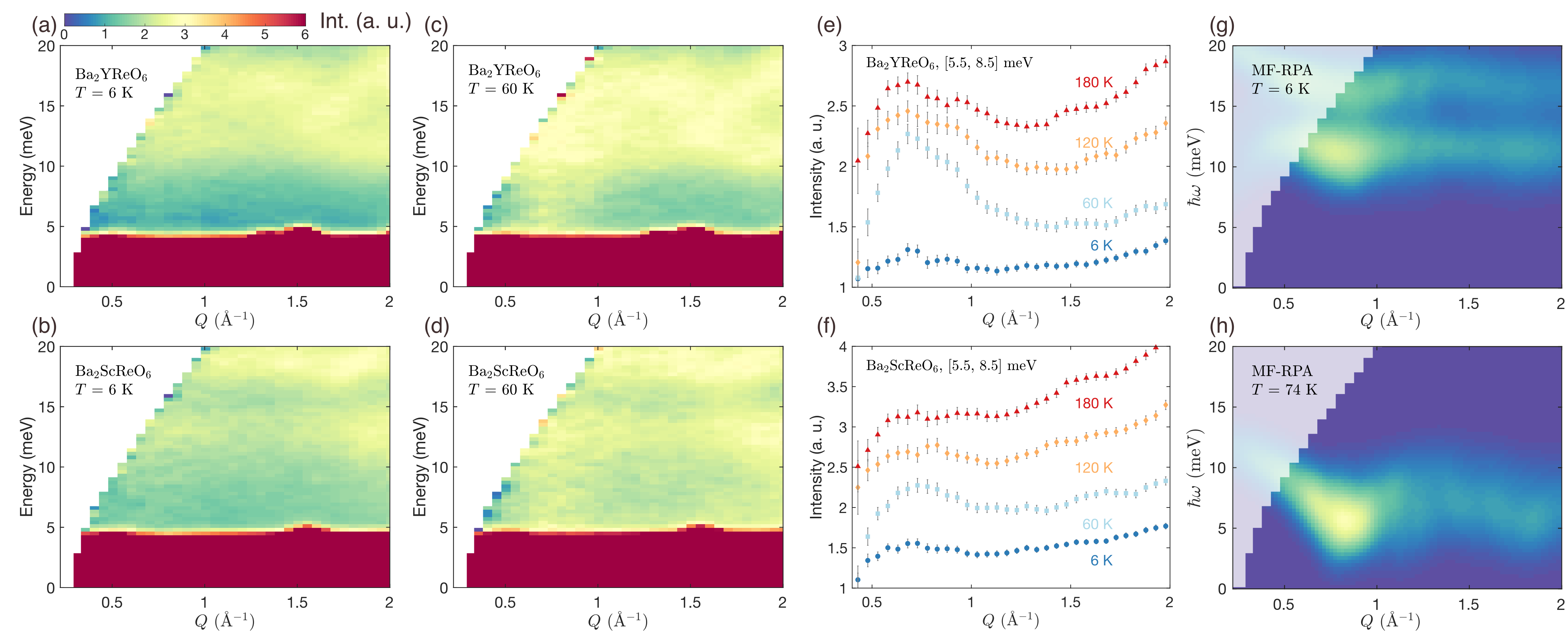}
    \caption{(a-d) INS spectra collected at $T=$ 6~K and 60~K for Ba$_2$YReO$_6$ and Ba$_2$ScReO$_6$ with incident energy $E_\mathrm{i}$ = 60~meV. (e, f) Temperature evolution of the integrated intensity in the energy range of [5.5, 8.5]~meV as a function of the wavevector transfer. (g, h) Simulated INS spectra using the MF-RPA method for the minimal model at $T=6$~K and 74~K. At low temperatures, the excitation is gapped by $\sim$~10~meV. At $T=74$~K, in the FQ phase, magnetic excitations are observed at $Q\sim$ 0.75~\AA$^{-1}$, which is consistent with the experimental observations at elevated temperatures.
    \label{fig:INS}}
\end{figure*}

\subsection{Phonon and magnon excitations}
Figures~\ref{fig:DFT}(a-d) present the INS data for Ba$_2$YReO$_6$ and Ba$_2$ScReO$_6$ collected on SEQUOIA at $T$ = 6~K with incident energies $E_\mathrm{i}$ = 60~meV and 225~meV. As can be inferred from the wavevector dependence, the phonon density of states dominate the spectra, overshadowing other possible contributions like magnetic excitations. Figures~\ref{fig:DFT}(e-h) present the theoretical phonon excitations calculated by the DFT, based on the lattice parameters as refined from the neutron diffraction data. The spectra have been convoluted by the instrumental energy resolution function. The calculated spectra reproduce the main features of the experimental results. However, in the low wavevector transfer regime ($Q < 2$~\AA$^{-1}$), between 10 and 20~meV, weak scattering is observed in experiments, which is not captured by the DFT calculations. These features may originate from magnetic excitations. Figures~\ref{fig:INS}(a-d) present the zoomed-in view of the INS spectra for Ba$_2$YReO$_6$ and Ba$_2$ScReO$_6$ collected at temperatures of $T$ = 6~K and $T$ = 60~K. Compared to the 6~K data, softened magnetic excitations at $T$ = 60~K can be observed at $Q\sim$ 0.75~\AA$^{-1}$, the same position for the magnetic Bragg peak at $(0, 0, 1)$, for energy transfers below 10~meV, a behavior that has been observed in other $5d$-based DPs~\cite{Morrow2016,Maharaj:PhysRevLett,
PhysRevB.98.104434,PhysRevB.91.075133,PhysRevB.93.220408}. Therefore, we conclude that at the base temperature of $T$ = 6~K, the magnetic scattering in the dipolar ordered phase is either out of the detectable area or overshadowed by the phonon excitations. Figures~\ref{fig:INS}(e) and~\ref{fig:INS}(f) illustrate the temperature evolution of the scattering intensity integrated in the range of [5.5, 8.5]~meV as a function of the wavevector transfer. At temperatures of 60, 120, and 180~K, the intensity at $Q \sim 0.75$~\AA$^{-1}$ becomes significantly enhanced, which strongly evidences the existence of softened magnetic excitations at temperatures above the dipolar long-range ordering transition.


\section{MF-RPA analysis}
Using the MF-RPA method, we propose a minimal model to describe the dipolar and quadrupolar correlations in Ba$_2$YReO$_6$ and Ba$_2$ScReO$_6$. In our model, we start from the single ion electronic configuration of the Re$^{5+}$ ions. The couplings among the Re$^{5+}$ ions include dominant AFM $J_1$ over the first neighbors, relatively weak ferromagnetic (FM) $J_2$ over the second neighbors, and quadrupolar interactions $K$ over the first neighbors. The exchange paths for the first- and second-neighbor couplings are indicated in Fig.~\ref{fig:crystal}.

The Hamiltonian of our minimal model is expressed as
\begin{equation}\mathcal{H}=\sum_i\mathcal{H}_{\mathrm{SI}}^i+\mathcal{H}_{\mathrm{I}},\end{equation}
where $\mathcal{H}_{\mathrm{SI}}^i$ represents the single ion Hamiltonian at site $i$ and $\mathcal{H}_{\mathrm{I}}$ describes the inter-ion interactions.

\subsection{Single ion Hamiltonian}
As discussed in the introduction section, the effective ground state of a Re$^{5+}$ ion in a strong octahedral crystal field is a SOC-induced $J_{\mathrm{eff}} = 2$ manifold. Within this manifold, the single ion Hamiltonian on site $i$ is expressed as
\begin{equation}
\mathcal{H}_{\mathrm{SI}}^i=\mathcal{H}_{\mathrm{RCF}}+\mathcal{H}_{\mathrm{W}}.
\end{equation} 
In this expression, $\mathcal{H}_{\mathrm{RCF}}$ represents the residual crystal field that splits the $J_{\mathrm{eff}}=2$ manifold into the $T_{2g}$ triplet and the $E_{g}$ doublet as shown in Fig.~\ref{fig:GS}. As discussed in Refs.~\cite{Maharaj:PhysRevLett,Paramekanti:PRB,Voleti:PRB}, this term arises from the higher-order perturbations and the nonspherical Coulomb interactions, which can be described as:
\begin{equation}
    \mathcal{H}_{\mathrm{RCF}}=-\frac{\Delta}{120}(\mathcal{O}_4^0+5\mathcal{O}_4^4),
\end{equation}
where the Steven operators are defined as:
\begin{equation}
\begin{aligned} 
\mathcal{O}_{4}^0&=35J_z^4-[30J(J+1)-25]J_z^2+3J^2(J+1)^2\\&-6J(J+1),\\\mathcal{O}_{4}^4&=\frac12(J_+^4+J_-^4).
\end{aligned}
\end{equation}
Here we choose negative $\Delta$ so that the energy of the triplet is lower than that of the doublet in the reversed level scheme as suggested for Ba$_2$YReO$_6$~\cite{Voleti:PRB}. For our semi-quantitative analysis, $\Delta$ is set to $-50$~meV based on the experimental Curie Weiss temperature. This choice is justified as the high Curie-Weiss temperature behavior is strongly impacted by the single ion physics~\cite{Voleti:PRB}. Adjustment of the exact value of $\Delta$ in a broad range of [$-$100, $-$35]~meV does not impact the conclusion of our analysis. We have also considered the case of $\Delta > 0$, but no satisfactory results were obtained. The second term in $\mathcal{H}_{\mathrm{SI}}$ represents a weak Weiss field, $\mathcal{H}_{\mathrm{W}}$, which accounts for the symmetry-broken term of the quadrupolar order. It is defined as
\begin{equation}
\mathcal{H}_{\mathrm{W}}= \lambda Q^{3z^2} =\lambda\left( 3J_z^2-J(J+1)\right)/\sqrt{3},
\end{equation}where $\lambda$ is a small constant that describes the strength of the Weiss field.

\subsection{Inter-ion interactions}
The exchange interactions among the Re$^{5+}$ ions are described by the inter-ion Hamiltonian $\mathcal{H}_{\mathrm{I}}$
\begin{equation}
\begin{aligned}
\mathcal{H}_{\mathrm{I}}&=\sum_{ij}\sum_{\alpha,\beta}\mathcal{J}_{\alpha\beta}(\Delta \mathbf{R}_{i,j})\mathcal{I}_\alpha^i\mathcal{I}_\beta^{j}\\&=\sum_{\langle i,j\rangle}(J_1\mathbf{J}_i\cdot\mathbf{J}_j+KQ^{3z^2,i}\cdot Q^{3z^2,j})+J_2\sum_{\langle\langle i,j\rangle\rangle}\mathbf{J}_i\cdot\mathbf{J}_j.
\end{aligned}
\end{equation}
where $\langle i,j\rangle$ and $\langle\langle i,j\rangle\rangle$ denote the first-neighbor (NN) and second-neighbor (NNN) interactions, respectively. $\mathcal{I}_\alpha^s$ are the multipolar tensor operators $\mathcal{I}=(J^x,J^y,J^z,Q^{3z^2},Q^{x^2-y^2})$ as defined in Refs.~\cite{Chen:PRB1,RevModPhys.81.807} within the $J_{\mathrm{eff}} = 2$ manifold. $\mathcal{J}_{\alpha\beta}(\Delta \mathbf{R}_{i,j})$ is the interaction matrix element between the $\alpha$ and $\beta$ operators where $\Delta \mathbf{R}_{i,j}=\mathbf{R}_{i}-\mathbf{R}_{j}$ is the vector connecting sites $i$ and $j$. In the matrix form, the couplings over the first and second neighbors can be expressed as
\begin{equation}
\mathcal J_{\mathrm{NN}} = \begin{pmatrix}
J_1 & 0 & 0 & 0 & 0\\
0 & J_1 & 0 & 0 & 0\\
0 & 0 & J_1 & 0 & 0\\
0 & 0 & 0 & K & 0\\
0 & 0 & 0 & 0 & 0\\
\end{pmatrix},
\end{equation}

\begin{equation}
\mathcal J_{\mathrm{NNN}} = \begin{pmatrix}
J_2 & 0 & 0 & 0 & 0\\
0 & J_2 & 0 & 0 & 0\\
0 & 0 & J_2 & 0 & 0\\
0 & 0 & 0 & 0 & 0\\
0 & 0 & 0 & 0 & 0\\
\end{pmatrix}.
\end{equation}

\subsection{Mean-field analysis of the ground state}
The interactions on a single ion can be approximated using the mean-field method. Following our experimental observation of a type-I dipolar AFM ground state, the mean-field Hamiltonian $\mathcal{H}_\mathrm{MF}^s$ can be defined on two sublattices as:
\begin{align}
    &\mathcal{H}_\mathrm{MF}^s=-\sum_{\alpha=1}^mH_\alpha^s\mathcal{I}_\alpha^s,\\
&H_\alpha^s=\sum_{s',\alpha}\mathcal{J}_{\alpha\beta}(\Delta \mathbf{R}_{s,s'})\langle\mathcal{I}_{\beta}^{s'}\rangle,
\end{align}
where $s$ represents different sublattices, and the summation includes all NN and NNN interactions. This mean-field Hamiltonian can be solved self-consistently to obtain the ordering behavior of both the dipolar and quadrupolar moments.

\subsection{Generalized susceptibility}
From the self-consistently determined mean-field ground state, one can calculate the single ion susceptibility for each sublattice $s$. According to the linear response theory,
\begin{equation}
    \small \chi_{\alpha\beta}^{0,s}(\omega)=\sum_{ij}\frac{\langle \Psi_i|\mathcal{I}_\alpha-\langle\mathcal{I}_\alpha\rangle|\Psi_j\rangle\langle \Psi_j|\mathcal{I}_\beta-\langle\mathcal{I}_\beta\rangle|\Psi_i\rangle}{\epsilon_{j}-\epsilon_i-\hbar\omega}(p_i-p_{j}),
\end{equation}
where $|\Psi_i\rangle$ and $\epsilon_i$ are the eigenstates and eigenvalues of the single ion Hamiltonian at site $s$, respectively. $p_i=\frac{\exp(-\epsilon_i/k_\mathrm{B}T)}{\sum_{j}\exp(-\epsilon_{j}/k_\mathrm{B}T)}$ is the thermal population factor with $k_\mathrm{B}$ denoting the Boltzmann's constant.

The generalized susceptibility can be calculated using the RPA method~\cite{Jensen:RareEarth,Rotter_2012,Boothroyd_neutronbook}. This method is widely used in the analysis of spin dynamics and has proven to be particularly successful in the study of multipolar interactions in the rare-earth systems~\cite{Jensen:RareEarth,Rotter_2012,PhysRevLett.105.167201,PhysRevB.84.104409,Leonid:PNAS}. Recently, it has also been applied to the study of the $5d$ heavy transition metal compounds~\cite{PhysRevLett.127.237201,PhysRevB.108.054436}. Under the RPA, the total generalized susceptibility is determined as, 
\begin{equation}
    \chi_{\alpha\beta}(\mathbf{Q},\omega)=\sum_{s,s'}\chi_{\alpha\beta}^{0,s}(\omega)\left[\delta^{ss'}-\chi_{\alpha\beta}^{0,s}(\omega)\mathcal{J}^{ss'}_{\alpha\beta}(\mathbf{Q})\right]^{-1},
    \label{eq:RPA}
\end{equation}
where $\delta^{ss'}$ is the Kronecker tensor, and $\mathcal{J}^{ss'}_{\alpha\beta}(\mathbf{Q})$ is the Fourier transform of the exchange interaction matrix element $\mathcal{J}_{\alpha\beta}(\mathbf{R}_{s}-\mathbf{R}_{s'})$.

The real ($\chi'(\mathbf{Q},\omega)$) and imaginary ($\chi''(\mathbf{Q},\omega)$) parts of the generalized susceptibility can be calculated separately following Eq.~\eqref{eq:RPA}. They are related by the Kramers-Kronig transformation~\cite{Boothroyd_neutronbook} as, 
\begin{equation}
    \chi'(\mathbf{Q},0)=\frac1\pi\int_{-\infty}^\infty\frac{\chi''(\mathbf{Q},\omega)}\omega \mathrm{d}\omega.
\end{equation}
The magnetic susceptibility as probed in our experiments can be calculated from the real part~\cite{Boothroyd_neutronbook} as,
\begin{align}
\chi_M &= \frac{V}{\mu_0}\chi'(0,0).
\end{align}
The INS cross section can be calculated from the imaginary part according to the fluctuation-dissipation theorem,
\begin{equation}
    \begin{aligned}
    S(\mathbf{Q},\omega)\propto\sum_{\alpha,\beta=1,2,3}(\delta_{\alpha\beta}-\hat Q_{\alpha}\hat Q_{\beta})
    \\\times\left[\sum_{{\mu,\mu^{\prime}}}F_{\alpha\mu}(\mathbf{Q})F_{{\beta\mu^{\prime}}}(\mathbf{Q}) \chi^{\prime\prime}_{{\mu\mu^{\prime}}}(\mathbf{Q},\omega)\right].
    \end{aligned}
\end{equation}
where $\mu,\mu'$ sum over the components of the multipolar tensor operators. $F(\mathbf{Q})$ is the form factor of the multipolar tensor operator. In our case, this form factor can be simplified to a dipolar form factor $F^2(Q)$ since the quadrupolar interaction does not contribute to the formfactor~\cite{PhysRevLett.127.237201},
\begin{equation}
    \begin{aligned}
    S(\mathbf{Q},\omega)\propto\sum_{\alpha,\beta=1,2,3}(\delta_{\alpha\beta}-\hat Q_{\alpha}\hat Q_{\beta})
    \\\times F^2(Q)\left[\sum_{{\alpha\beta}}\chi^{\prime\prime}_{{\alpha\beta}}(\mathbf{Q},\omega)\right].
    \end{aligned}
\end{equation}
\subsection{Applications to Ba$_2$YReO$_6$ and Ba$_2$ScReO$_6$}
\begin{figure}[htb]
    \includegraphics[width=0.47\textwidth]{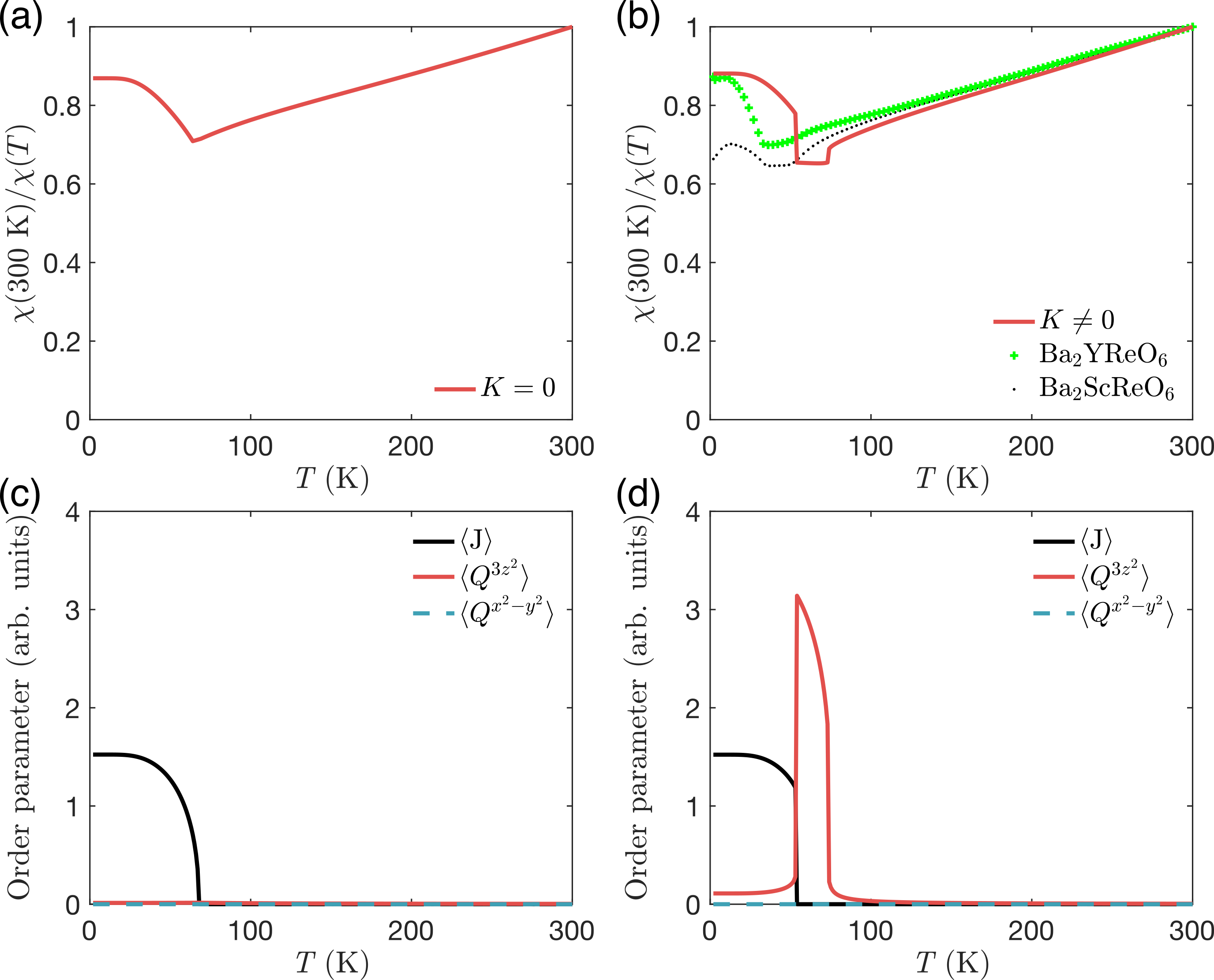}
    \caption{(a-b) Temperature dependence of the normalized inverse magnetic susceptibility, $\chi(300\mathrm{K})/\chi(T)$, where $\chi(300\mathrm{K})$ is the susceptibility at $T$ = 300~K. Red solid line represents the calculation. Green crosses and black dots reproduce the experimental results for  Ba$_2$YReO$_6$ and Ba$_2$ScReO$_6$ as shown in Fig.~\ref{fig:chi}. In panel (a), a model with $K=0$ shows only dipolar AFM transition. In panel (b), the minimal model including quadrupolar interaction $K$ reproduces the experimentally observed plateau. Temperature dependence of the order parameters for (c) the model with $K=0$, and for (d) the minimal model.
    \label{fig:sus}}
\end{figure}

Figure~\ref{fig:sus} presents the calculated temperature dependence of the inverse magnetic susceptibility $\chi(300\mathrm{K})/\chi(T)$ and the order parameters using the MF-RPA method. The order parameters are defined as,
\begin{align}
\langle \mathrm{J}\rangle&=\frac12\left\lvert\langle \mathbf{J}^1\rangle-\langle \mathbf{J}^2\rangle\right\rvert\\
\langle Q^{3z^2}\rangle &= \frac12\left\lvert\langle Q^{3z^2,1}\rangle+\langle Q^{3z^2,2}\rangle\right\rvert\\
\langle Q^{x^2-y^2}\rangle &= \frac12\left\lvert\langle Q^{x^2-y^2,1}\rangle+\langle Q^{x^2-y^2,2}\rangle\right\rvert
\end{align}
where the last index in the superscript represents the sublattices 1 and 2. The parameters for the calculations, including the coupling strengths, are listed in Table~\ref{table:param}. The magnetic susceptibility and order parameters for the minimal model in Fig.~\ref{fig:sus}(b) exhibit two transitions, consistent with our experimental observations for Ba$_2$YReO$_6$ and Ba$_2$ScReO$_6$. In contrast, models with quadrupolar interaction $K=0$ in Fig.~\ref{fig:sus}(a) show only one transition that corresponds to the dipolar order. Compared to the experiments, the transition temperatures are overestimated, possibly due to the limitation of the mean-field approximations. At 100~K, the quadrupolar moment $Q^{3z^2}$ starts to develop a ferroic order as shown in Fig.~\ref{fig:sus}(d), causing the magnetic susceptibility to deviate from the Curie-Weiss behavior. Between 53 and 75~K, $\chi(300\mathrm{K})/\chi(T)$ form a plateau, similar to the experimental results in Fig.~\ref{fig:chi}. This intermediate phase is associated with the FQ order as evidenced by the finite quadrupolar moment $Q^{3z^2}$ in Fig.~\ref{fig:sus}(d), which is triggered by the quadrupolar interactions. This phase does not appear in the $K=0$ case as illustrated in Fig.~\ref{fig:sus}(c). At 53~K, the system transitions into a type-I dipolar AFM state as indicated by the evolution of the order parameter $\langle \mathrm{J}\rangle$. The quadrupolar moment remains ferroic ordered even in the AFM phase at low temperatures in our minimal model, although its ordered moment size becomes greatly reduced due to the competition with the dipolar order.

The calculated INS spectra for the same minimal model, parameters from Table~\ref{table:param}, are shown in Figs.~\ref{fig:INS}(g) and~\ref{fig:INS}(h). At $T = 6$~K, the magnetic excitations are gapped, which is consistent with the experiments. In the FQ state at $T = 74$~K, the excitation band shifts towards the elastic line, leading to low-energy excitations that are very similar to the experimental INS spectra as shown in Figs.~\ref{fig:INS}(c) and~\ref{fig:INS}(d). At higher temperatures, the experimental INS spectra are determined by short-range correlations that are not captured in our mean-field analysis.
\begin{table}[h!]
\caption{Main parameters for the minimal model. All parameters are shown in units of meV.}
\begin{ruledtabular}
\begin{tabular}{cccccc}
\hline
Parameter&$\Delta$ & $J_1$ & $J_2=0.2J_1$ &$K=-0.07J_1$&$\lambda$\\\hline
Value&$-50$ & 1.15 & $-0.23$ & $-0.0805$ & $-0.015$
\label{table:param}
\end{tabular}
\end{ruledtabular}
\end{table}

It is noteworthy that the actual interactions in Ba$_2$YReO$_6$ and Ba$_2$ScReO$_6$ can be more complicated than those considered in our minimal model. For instance, in systems with strong SOC, anisotropic interactions between the multipolar moments may play a significant role~\cite{PhysRevLett.127.237201,PhysRevB.108.054436,Leonid:PNAS}. Refinement of a complete set of the parameters described in our Hamiltonian is beyond the quality of the data. The phonon contribution to the scattering and the powder averaged spectrum only allow for a qualitative comparison of these terms. However, our minimal model is able to reproduce the main features of the experimental results and provides a good starting point for further studies on the multipolar correlations in the $5d^2$ double perovskites.

\section{Conclusion}
We have performed magnetic susceptibility, heat capacity, and neutron scattering experiments on the $5d^2$ rhenium-based DP compounds Ba$_2$YReO$_6$ and Ba$_2$ScReO$_6$ to study their dipolar and quadrupolar correlations. Both compounds exhibit a type-I dipolar AFM ground state with a propagation vector $\mathbf{q} = (0, 0, 1)$, where the ordered moment size is significantly reduced due to SOC and quadrupolar correlations. Our INS experiments reveal dominant phonon excitations, which are well described by the DFT calculations. Weak magnetic excitations with an excitation gap are observed at low temperatures, which become softened at elevated temperatures. Using the MF-RPA method, a minimal model that considers both the dipolar and quadrupolar interactions is proposed to explain the susceptibility data and INS spectra, which evidences the existence of a FQ order in both the paramagnetic phase and the AFM ordered ground state.

\begin{acknowledgments}
This work was supported by the U.S. Department of Energy, Office of Science, Basic Energy Sciences, Materials Sciences and Engineering Division (Neutron scattering data collection and initial analysis). This research used resources at the Spallation Neutron Source (SNS) and the High Flux Isotope Reactor (HFIR), both of which are DOE Office of Science User Facilities operated by the Oak Ridge National Laboratory (ORNL). The beam time was allocated to POWGEN on proposal number IPTS-27345, to SEQUOIA on proposal number IPTS-26621, and to HB-1A on proposal number IPTS-26600. Works at USTC were funded by the National Key R\&D Program of China under the Grant No. 2024YFA1613100 and the National Natural Science Foundation of China (NSFC) under the Grant No. 12374152. This work was also supported by the Japan Society for the Promotion of Science KAKENHI under the Grant No. JP20H01858, No. JP23H04860, and No. JP24H01187, and the JSPS Bilateral Open Partnership Joint Research Projects JPJSBP120239915.
\end{acknowledgments}

%

\end{document}